# COMPARATIVE SURVIVAL ANALYSIS OF *DEINOCOCCUS RADIODURANS* AND THE HALOARCHAEA *NATRIALBA MAGADII* AND *HALOFERAX VOLCANII*, EXPOSED TO VACUUM ULTRAVIOLET IRRADIATION

Ximena C. Abrevaya[1]; Ivan G. Paulino-Lima[2], Douglas Galante[3], Fabio Rodrigues[4], Pablo J.D. Mauas[1,5], Eduardo Cortón[6], and Claudia de Alencar Santos Lage[2]

[1]Instituto de Astronomía y Física del Espacio (IAFE), Universidad de Buenos Aires-CONICET. Buenos Aires, Argentina.

[2]Instituto de Biofísica Carlos Chagas Filho, Universidade Federal do Rio de Janeiro. Rio de Janeiro, Brazil.

[3]Instituto de Astronomia, Geofísica e Ciências Atmosféricas, Universidade de São Paulo. São Paulo, Brazil.

[4]Instituto de Química, Universidade de São Paulo. São Paulo, Brazil.

[5]Facultad de Ciencias Exactas y Naturales, Universidad de Buenos Aires. Buenos Aires, Argentina.

[6]Grupo de Biosensores y Bioanálisis, Departamento de Química Biológica, Facultad de Ciencias Exactas y Naturales, UBA and CONICET. Buenos Aires, Argentina.

Correspondence: Ximena C. Abrevaya. Address: Pabellón IAFE. CC 67, Suc. 28, 1428. Buenos Aires, Argentina. Phone: (+54)-11-4789-0179 ext.: 105. Fax: (+54)-11-4786-8114. E-mail: abrevaya@iafe.uba.ar

# ABSTRACT

The haloarchaea *Natrialba magadii* and *Haloferax volcanii*, as well as the radiation-resistant bacterium *Deinococcus radiodurans*, were exposed to vacuum-UV (V-UV) radiation at the Brazilian Synchrotron Light Laboratory (LNLS). Cell monolayers (**containing $10^5 - 10^6$ cells per sample**) were prepared over polycarbonate filters and irradiated under high vacuum ($10^{-5}$ Pa) with polychromatic synchrotron radiation. *N. magadii* was remarkably resistant to high vacuum with a survival fraction of (($3.77 \pm 0.76$) x $10^{-2}$), larger than the one of *D. radiodurans* (($1.13 \pm 0.23$) x $10^{-2}$). The survival fraction of the haloarchaea *H. volcanii,* of (($3.60 \pm 1.80$) x $10^{-4}$), was much smaller. Radiation resistance profiles were similar between the haloarchaea and *D. radiodurans* for fluencies up to 150 J m$^{-2}$. For fluencies larger than 150 J m$^{-2}$ there was a significant decrease in the survival of haloarchaea, and in particular *H. volcanii* did not survive. Survival for *D. radiodurans* was 1% after exposure to the higher V-UV fluency (1350 J m$^{-2}$) while *N. magadii* had a survival lower than 0.1%. Such survival fractions are discussed regarding the possibility of interplanetary transfer of viable micro-organisms and the possible existence of microbial life in extraterrestrial salty environments such as the planet Mars and the Jupiter's moon Europa. This is the first work reporting survival of haloarchaea under simulated interplanetary conditions.

**INTRODUCTION**

The study of microbial ability to survive in space conditions has multiple applications in Astrobiology. For example, it is important to develop planetary protection procedures (Raulin et al., 2010), life support systems (Hendrickx and Mergeay, 2007) and energy fuel cells based on a number of microbial species (Flinn, 2004). It is also important to avoid forward contamination (DeVincenzi et al., 1998; Crawford, 2005) which, among other things, can compromise *in situ* detection of life outside Earth (see Abrevaya et al., 2010 and references therein). This problem is also relevant to the Panspermia hypothesis (Arrhenius, 1903), which proposes that life can spread through space.

The extraterrestrial environment is considered lethal for organisms, due to the high levels of radiation, high vacuum conditions and extreme temperatures, which could have impact over relevant macromolecules as nucleic acids, leading to increased mutation rates, cell damage and inactivation (Horneck, 1999; Paulino-Lima et al., 2010). In particular, UV radiation has been mostly studied because it causes extensive damage on cells and can determine the conditions of habitability elsewhere (see, for example, Buccino et al., 2006, 2007).

Survival of microorganisms and biomolecules under space conditions has been evaluated in many ways, either under direct exposure in real flight missions around Earth's orbit or under simulated space UV radiation and/or vacuum (Horneck et al., 2010; Olsson-Francis and Cockell, 2010). Different microorganisms have been tested, including bacteria, fungi, bacterial or fungal spores and viruses, as well as biomolecules such as DNA, amino acids and liposomes (Olsson-Francis and Cockell, 2010; Horneck et al., 2010).

Negligible UV radiation shorter than **290** nm reaches the Earth's surface, since it is strongly absorbed by the atmosphere, but it should be considered in studies concerning life in extraterrestrial environments. Unfortunately, the experimental techniques to work with these wavelengths are complex, due to the requirement of vacuum conditions and high radiation fluxes (Cefalas et al., 1993; Sarantopoulou et al., 1994). Nonetheless, a number of authors report effects of such wavelengths in microorganisms (Koike and Oshima, 1993; Cefalas et al., 2001; Saffary et al., 2002; Schuerger et al., 2003; Heise et al., 2004; Newcombe et al., 2005; Clauss, 2006; Sarantopoulou et al., 2006a; Schuerger et al., 2006; Tauscher et al., 2006; Fajardo-Cavazos et al., 2010; Wassmann et al., 2010; Galletta et al., 2010, Sarantopoulou et al., 2011).

In the present study we investigate the survival of two different non-sporulating halophilic archaea or haloarchaea (family *Halobacteriaceae*), *Haloferax volcanii* (Mullakhanbhai and Larsen, 1975) and *Natrialba magadii* (Tindall et al., 1984; Kamekura et al., 1997) for the first time in a extraterrestrial simulation facility. It is well known that extreme halophiles as haloarchaea are relevant for Astrobiology, due to their capacity not only to dwell on environments of high salinity, e.g: Mars (Mancinelli et al., 2004) or Jupiter's moon Europa (Marion et al., 2003), but also for their ability to cope with extreme temperatures, pH and radiation.

These strains, as well as the radiation resistant non-sporulating bacterium *Deinococcus radiodurans,* were desiccated and subjected to UV and vacuum conditions similar to those found in interplanetary space near Earth orbit. Our experiments took place at the Brazilian Synchrotron Light Laboratory (LNLS) located in Campinas, Brazil, using the synchrotron beamline equipped with a Toroidal Grating Monochromator (TGM) and an end station with a vacuum chamber (Cavasso et al., 2007).

**MATERIAL AND METHODS**

*H. volcanii* (DS70) and *N. magadii* (ATCC 43099) strains were kindly provided by Dr. R. E. de Castro, Universidad Nacional de Mar del Plata,

Argentina, and *D. radiodurans* (R1 wild type strain) was obtained at Instituto de Radioproteção e Dosimetria, Rio de Janeiro, Brazil.

*H. volcanii* was grown aerobically at 30ºC with shaking at 200 rpm and was cultivated in Hv-YPC broth (Kauri et al., 1990) containing (g $l^{-1}$): yeast extract (5), peptone (1), casaminoacids (1), NaCl (144), $MgSO_4.7H_2O$ (21), $MgCl.6H_2O$ (18), KCl (4.2), $CaCl_2$ (3mM), and Tris-HCl (12mM), with pH adjusted to 6.8. *N. magadii* was grown aerobically at 37°C with shaking at 200 rpm. Growth medium composition was (g $l^{-1}$): yeast extract (5); NaCl (200); $Na_2CO_3$ (18.5); sodium citrate (3); KCl (2); $MgSO_4.7H_2O$ (1); $MnCl_2.4H_2O$ (3.6 x $10^{-4}$); $FeSO_4.7H_2O$ (5 x $10^{-3}$), with pH adjusted to 10 (modified from Tindall et al., 1984). *D. radiodurans* was cultivated in TGY broth containing (g $l^{-1}$) tryptone (10); yeast extract (6); glucose (2) (Anderson et al., 1956) with shaking at 200 rpm, at 32ºC for 10h, or until the early stationary phase was reached.

The optical density of the cultures was spectrophotometrically measured at λ=600 nm and aliquots of 10 µl were taken for direct cell counting under the microscope. Results were compared to colony forming units (CFU) grown on agar-solidified culture media in order to assess the viability of the culture. After the estimation of the cell concentration, 1 ml of haloarchaea cultures was washed twice in saline solutions with the same composition and pH of the culture medium, but without organic compounds.

Similarly, 1 ml of the *D. radiodurans* culture was washed twice in previously-sterilized distilled water. At this stage, aliquots of these cultures were separated and kept outside the vacuum chamber to be used as controls. Other aliquots were taken to expose the cells to vacuum alone and to vacuum plus UV in the TGM workstation. All experiments were performed in triplicate.

The samples to be exposed were prepared using a hexagonal copper-made sample-holder that can be introduced inside the vacuum chamber of the workstation. Several square pieces of polycarbonate filter (Millipore) with a surface area of approximately 25 mm$^2$ were mounted on the surface of the sample-holder using double-sided carbon tape. The system was sterilized by exposure to germicidal UV lamp (254nm, 1620 J m$^{-2}$) in a laminar flow. After sterilization, a volume of 1 µl of the cell suspensions (containing $10^8 – 10^9$ cells per ml) was loaded on the polycarbonate filters (monolayers containing $10^5 – 10^6$ cells per sample). The system was kept inside the laminar flow for at least 30 minutes for dehydration of the samples. The sample-holder was then placed inside the vacuum chamber at the TGM beamline workstation.

Inside this chamber, cells were exposed to decreasing vacuum pressure along 3 hours, due to the required pumping time needed to reach high vacuum ($10^{-5}$ Pa) from atmospheric pressure, before irradiation. This pressure was kept during the whole irradiation experiment, which was around 1 hour. Finally, it took 15 minutes to vent the chamber.

Samples were exposed at room temperature to UV plus vacuum UV (V-UV) radiation. A gas filter (neon) was interposed between the beamline and the experimental chamber to attenuate the X-ray portion of the synchrotron radiation spectrum, below 57.6 nm. The synchrotron emission reached 400 nm, but the emitted energy decays exponentially beyond this, resulting in an effective cut-off of longer wavelengths after 124 nm. Therefore, the complete wavelength ranged from 57.6 to 124 nm (Figure 1). The exposure times were 0, 10, 30, 100, 300 and 900 seconds, resulting in fluencies of 0, 15, 45, 150, 450 and 1350 $Jm^{-2}$, respectively. The photon flux was measured using photodiodes to check for electrical current variations in real-time. The irradiance of the TGM beamline is compared with those at the orbits of the Earth, Mars and Europa in Table 2.

After exposure, 10 μl of culture medium were dropped on the polycarbonate filters and mixed with a micropipette for 10 seconds to recover the cells. Afterwards, samples were diluted and aliquots seeded on nutrient plates which were incubated. The efficiency of cell recovery was determined to be between 40 and 50% by comparison with cell counts scored in non-deposited samples.

Survival fractions were calculated as $N/N_0$, where **N** is the number of colony-forming units (CFUs) after treatment and $N_0$ is the number of CFUs corresponding to counts of the non-irradiated sample. Survival fractions for the exposure to vacuum alone and for the exposure to vacuum plus irradiation at several doses are shown in Figures 2 and 3, respectively.

To assess the statistical significance of the results, ANOVA tests were performed, followed by a Fisher´s least difference mean separation test (p=0.05) using the OriginPro8 software (OriginLab corporation, USA). Since the difference between our data is close to, or higher than one order of magnitude, data were log-transformed to achieve homogeneity of variances.

**RESULTS**

The effects of vacuum alone on survival fractions of *N. magadii, H. volcanii* and *D. radiodurans* were measured. As shown in Fig. 2, the haloarchaea *N. magadii* showed the maximum survival fraction to high vacuum ($10^{-5}$ Pa), which corresponds to $(3.77 \pm 0.76) \times 10^{-2}$. This fraction was almost 3-fold higher than for *D. radiodurans,* which had a survival fraction of $(1.13 \pm 0.23) \times 10^{-2}$ under the same conditions. Surprisingly, the survival fraction of the haloarchaea *H. volcanii* was remarkably low compared with *N. magadii* and *D. radiodurans*, being $(3.60 \pm 1.80) \times 10^{-4}$, which is two orders of magnitude below the values obtained for the other strains. The statistical analysis showed that the survival fractions are significantly different at p<0.05 in all cases.

The samples that were subjected to vacuum only were kept in the same chamber where the irradiations were performed. Therefore, the values obtained for the vacuum exposure alone were taken as controls in order to calculate the survival fractions obtained for V-UV experiments.

While being exposed to high vacuum, both haloarchaea and *D. radiodurans* were exposed to synchrotron V-UV radiation at different doses (0, 15, 45, 150, 450 and 1350 Jm$^{-2}$). As shown in Figure 3, for fluencies up to 150 Jm$^{-2}$ the three strains show similar survival fractions. In fact, the differences in the survival fractions up to 150 Jm$^{-2}$ are not significant according to the statistical tests ($p>0.05$). At higher fluencies, above 150 Jm$^{-2}$, the survival of the haloarchaeal strains diminishes much more rapidly in comparison to *D. radiodurans*. In particular, we detected no survivors for *H. volcanii*. Statistical analysis showed that these differences are significant ($p<0.05$). Fluencies resulting in 10% of survivors in the cell population (LD$_{90}$) were similar between haloarchaea and *D. radiodurans*, between 200 and 300 Jm$^{-2}$.

**DISCUSSION**

It is known that *D. radiodurans* is very resistant to desiccation under the present simulation setup (Paulino-Lima et al., 2010). For example, in an experiment using simulated Martian soil under low pressure, 30% of the initial community of *D. radiodurans* survived for 10 days while *E. coli* did not survive (Diaz and Schulze-Makuch, 2006). However, our results show that *N.*

*magadii* has a survival fraction three-fold higher than that for *D. radiodurans* (Figure 2). This could be explained by a more efficient mechanism against desiccation present in haloarchaea in general, due to their their adaption to environmental stresses caused by the high temperatures and salt concentrations in their natural habitat. A particular osmoadaptation mechanism in haloarchaea is based on the presence of high levels of intracellular concentrations of $K^+$ ions inside the cell (Oren, 1999). For example, Kottemann et al. (2005) found that 25% of *Halobacterium* cells survived after 20 days under high vacuum ($10^{-6}$ Pa).

On the other hand, exposure of *H. volcanii* to vacuum reduced its survival 100-fold compared to that of *N. magadii*. It should be noted that the protection mechanisms against desiccation should be stronger for *N. magadii*, which is an extreme halophile living in salt concentrations between 3.5-4.0 M NaCl, than for *H. volcanii*, a moderate halophile which lives in 1.7-2.5 M NaCl range. Furthermore, in special conditions, such as nutrient starvation, haloalkalophilic strains like *N. magadii* are also capable to produce an organic solute (2-sulfotrehalose), which in part replaces intracellular KCl (Desmarais et al., 1997), a mechanism which is probably not present in *H. volcanii*, which lives in neutral pH.

Regarding UV resistance, McCready et al. (2005) compared the resistance profiles to UV-C between the haloarchaea *Halobacterium* NRC-1

and *D. radiodurans* at different fluencies (between 0 and 200 Jm$^{-2}$) showing that both microorganisms are highly resistant to UV-C radiation.

In this paper, the V–UV range from 57.6 to 124 nm was focused as astrobiologically relevant in terms of microbial life spreading within the Solar System. Interestingly, unlike UV-C photons, which penetrate deep into the cell and damage DNA, V-UV photons can be mostly absorbed by the cell membrane (Cefalas, 2005; Sarantopoulou et al., 2006). In particular, Yagi et al. (2009) found that at 62 nm the absorption depth is of around 1 nm, much thinner than a typical haloarchaea membrane, of tens of nm depending on the composition of the membrane (Steensland and Larsen, 1969). All the microorganisms studied in this work have similar cell walls, since *D. radiodurans* has an atypical wall of similar composition to haloarchaea. Consistently, our radiation resistance profiles were similar between both haloarchaea and *D. radiodurans*, at least for fluencies up to 150 Jm$^{-2}$ (Figure 3).

For fluencies greater than 150 Jm$^{-2}$, survival fractions are higher for *D. radiodurans* than for haloarchaea. In particular, *H. volcanii* did not survive beyond that fluency, probably due to synergistic effects between V-UV and high vacuum. It should be noted that *D. radiodurans* survival fraction is two-fold higher than *N. magadii*, and therefore the radiation pathway to internal components of the cell as genetic material is twice as large. We point out that, although the composition of the cell wall is similar between *D. radiodurans* and haloarchaea, the cell wall structure of *N. magadii* is not known, but should account for the observed V-UV resistance.

**Conclusions**

This is the first work reporting survival of haloarchaea under simulated interplanetary conditions. We measured the survival of haloarchaeal cells *Natrialba magadii* and *Haloferax volcanii* under irradiation with V-UV photons on the range 57.5-124 nm. We used a V-UV flux similar to the solar one at 1 astronomical unit and pressures similar to those found in low Earth orbit or on the surface of Europa (which is below $10^{-4}$ Pa). The survival curves were compared with the response of *D. radiodurans*, a microorganism considered a good candidate to endure the extreme conditions found in space and on the surface of other planets and moons (Clauss, 2006; de La Vega et al., 2007; Diaz and Schulze-Makuch, 2006; Dose et al., 1996; Saffary et al., 2002; Paulino-Lima et al., 2010).

Our results indicate that unprotected *D. radiodurans* cells drop to 1% survival at 1350 Jm$^{-2}$, while *N. magadii* survives some 0.1% to the same exposure. *H. volcanii*, on the other hand, did not survive beyond 150 Jm$^{-2}$ fluencies. This survival fractions show that cells of *H. volcanii* and *N. magadii* fully exposed to solar vacuum-UV irradiation on a planetary surface or on meteorites would be quickly depleted by at least three orders of magnitude. However, several cells did survive, and much longer exposure times need to be tested to see if at least a small number of cells of *N. magadii* and *D. radiodurans* could survive the V-UV and vacuum damages present in space without any protection.

**Acknowledgements**

We would like to thank Laboratorio Nacional de Luz Síncrotron (LNLS), and Centro de Ciência e Tecnologia do Bioetanol, Campinas, Brazil, which partially supported this work and also provided the facility and the staff supervision to conduct part of the experiments. We acknowledge the comments by two referees, which helped to improve the paper.